# Application of Generalized Quantum Hydrodynamics In the Theory of Quantum Soliton's Evolution, Atom Structure and Lightning Ball

Boris V. Alexeev

Moscow Academy of Fine Chemical Technology (MITHT)
Prospekt Vernadskogo, 86, Moscow 119570, Russia
B.Alexeev@ru.net

**Abstract**

Quantum solitons are discovered with the help of generalized quantum hydrodynamics (GQH). The solitons have the character of the stable quantum objects in the self consistent electric field. These effects can be considered as explanation of the existence of lightning balls. The delivered theory demonstrates the great possibilities of the generalized quantum hydrodynamics in investigation of the quantum solitons. The paper can be considered also as comments and prolongation of the materials published in the known author`s monograph (Boris V. Alexeev, Generalized Boltzmann Physical Kinetics. Elsevier. 2004). The theory leads to solitons as typical formations in the generalized quantum hydrodynamics.

## 1. Introduction. About the basic principles of the Generalized Quantum Hydrodynamics (GHD).

I begin with the short reminding of basic principles of GQH created in particular in [1- 3]. As it is shown (see, for also [4]) the theory of transport processes (including quantum mechanics) can be considered in the frame of unified theory based on the non-local physical description. In particular the generalized hydrodynamic equations represent an effective tool for solving problems in the very vast area of physical problems. For simplicity in introduction, we will consider fundamental methodic aspects from the qualitative standpoint of view avoiding excessively cumbersome formulas. A rigorous description is found, for example, in the monograph [4].

Transport processes in open dissipative systems are considered in physical kinetics. Therefore, the kinetic description is inevitably related to the system diagnostics. Such an element of diagnostics in the case of theoretical description in physical kinetics is the concept of the physically infinitely small volume (**PhSV**). The correlation between theoretical description and system diagnostics is well-known in physics. Suffice it to recall the part played by test charge in electrostatics or by test circuit in the physics of magnetic phenomena. The traditional definition of **PhSV** contains the statement to the effect that the **PhSV** contains a sufficient number of particles for introducing a statistical description; however, at the same time, the **PhSV** is much smaller than the volume $V$ of the physical system under consideration; in a first approximation, this leads to local approach in investigating the transport processes. It is assumed in classical hydrodynamics that local thermodynamic equilibrium is first established within the **PhSV**, and only after that the transition occurs to global thermodynamic equilibrium if it is at all possible for the system under study. Let us consider the hydrodynamic description in more detail from this point of view. Assume that we have two neighboring physically infinitely small volumes **PhSV$_1$** and **PhSV$_2$** in a nonequilibrium system. The one-particle distribution function (DF) $f_{sm,1}(\mathbf{r}_1, \mathbf{v}, t)$ corresponds to the volume **PhSV$_1$**, and the function $f_{sm,2}(\mathbf{r}_2, \mathbf{v}, t)$ — to the volume **PhSV$_2$**. It is assumed in a first approximation that $f_{sm,1}(\mathbf{r}_1, \mathbf{v}, t)$ does not vary within **PhSV$_1$**, same as $f_{sm,2}(\mathbf{r}_2, \mathbf{v}, t)$ does not vary within the neighboring volume **PhSV$_2$**. It is this assumption of locality that is implicitly contained in the Boltzmann equation (BE). However, the assumption is too crude. Indeed, a particle on the boundary between two volumes, which experienced the last collision in **PhSV$_1$** and moves toward **PhSV$_2$**, introduces information about the $f_{sm,1}(\mathbf{r}_1, \mathbf{v}, t)$ into the neighboring volume **PhSV$_2$**. Similarly, a particle on the boundary between two volumes, which experienced the last collision in **PhSV$_2$** and moves toward **PhSV$_1$**, introduces information about the DF $f_{sm,2}(\mathbf{r}_2, \mathbf{v}, t)$ into the neighboring volume **PhSV$_1$**. The relaxation over translational degrees of freedom of particles of like masses occurs during several collisions. As a result, "Knudsen layers" are formed on the boundary between neighboring physically infinitely small volumes, the characteristic dimension of which is of the order of path length. Therefore, a correction must be introduced into the DF in the **PhSV**, which is proportional to the mean time between collisions and to the substantive derivative of the DF being measured (rigorous derivation is given in [4]). Let a particle of finite radius be characterized as before by the position **r** at the instant of time *t* of its center of mass moving at velocity **v**. Then, the situation is possible where, at some instant of time *t*, the particle is located on the interface between two volumes. In so doing, the lead effect is possible (say, for **PhSV$_2$**), when the center of mass of particle moving to the neighboring volume **PhSV$_2$** is still in **PhSV$_1$**. However, the delay effect takes place as well, when the center of mass of particle moving to the neighboring volume (say, **PhSV$_2$**) is already located in **PhSV$_2$** but a part of the particle still belongs to **PhSV$_1$**. This entire complex of effects defines non-local effects in space and time.

The physically infinitely small volume (**PhSV**) is an *open* thermodynamic system *for any division of macroscopic system by a set of PhSVs.* However, the BE [3, 4]

$$Df/Dt = J^B, \qquad (1.1)$$

where $J^B$ is the Boltzmann collision integral and $D/Dt$ is a substantive derivative, fully ignores non-local effects and contains only the local collision integral $J^B$. The foregoing nonlocal effects are insignificant only in equilibrium systems, where the kinetic approach changes to methods of statistical mechanics.

This is what the difficulties of classical Boltzmann physical kinetics arise from. Also a weak point of the classical Boltzmann kinetic theory is the treatment of the dynamic properties of interacting particles. On the one hand, as follows from the so-called "physical" derivation of the BE, Boltzmann particles are regarded as material points; on the other hand, the collision integral in the BE leads to the emergence of collision cross sections.

The rigorous approach to derivation of kinetic equation relative to one-particle DF *f* ($KE_f$) is based on employing the hierarchy of Bogoliubov equations. Generally speaking, the structure of $KE_f$ is as follows:

$$\frac{Df}{Dt} = J^B + J^{nl}, \qquad (1.2)$$

where $J^{nl}$ is the non-local integral term. An approximation for the second collision integral is suggested by me in *generalized* Boltzmann physical kinetics,

$$J^{nl} = \frac{D}{Dt}\left(\tau \frac{Df}{Dt}\right). \tag{1.3}$$

Here, $\tau$ is the mean time *between* collisions of particles, which is related in a hydrodynamic approximation with dynamical viscosity $\mu$ and pressure $p$,

$$\tau\, p = \Pi \mu, \tag{1.4}$$

where the factor $\Pi$ is defined by the model of collision of particles: for neutral hard-sphere gas, $\Pi = 0.8$ [5]. All of the known methods of deriving kinetic equation relative to one-particle DF lead to approximation (1.3), including the method of many scales, the method of correlation functions, and the iteration method.

Fluctuation effects occur in any open thermodynamic system bounded by a control surface transparent to particles. GBE (1.2) leads to generalized hydrodynamic equations [4] as the local approximation of non local effects, for example, to the continuity equation

$$\frac{\partial \rho^a}{\partial t} + \frac{\partial}{\partial \mathbf{r}} \cdot (\rho \mathbf{v}_0)^a = 0, \tag{1.5}$$

where $\rho^a$, $\mathbf{v}_0^a$, $(\rho \mathbf{v}_0)^a$ are calculated in view of non-locality effect in terms of gas density $\rho$, hydrodynamic velocity of flow $\mathbf{v}_0$, and density of momentum flux $\mathbf{v}_0$; for locally Maxwellian distribution, $\rho^a$, $(\rho \mathbf{v}_0)^a$ are defined by the relations

$$(\rho - \rho^a)/\tau = \frac{\partial \rho}{\partial t} + \frac{\partial}{\partial \mathbf{r}} \cdot (\rho \mathbf{v}_0), \quad (\rho \mathbf{v}_0 - (\rho \mathbf{v}_0)^a)/\tau = \frac{\partial}{\partial t}(\rho \mathbf{v}_0) + \frac{\partial}{\partial \mathbf{r}} \cdot \rho \mathbf{v}_0 \mathbf{v}_0 + \vec{I} \cdot \frac{\partial p}{\partial \mathbf{r}} - \rho \mathbf{a}, \tag{1.6}$$

where $\vec{I}$ is a unit tensor, and $\mathbf{a}$ is the acceleration due to the effect of mass forces.

In the general case, the parameter $\tau$ is the non-locality parameter; in quantum hydrodynamics, its magnitude is defined by the "time-energy" uncertainty relation [2]. The violation of Bell's inequalities [6, 7] is found for local statistical theories, and the transition to non-local description is inevitable.

The following conclusion of principal significance can be done from the previous consideration [1, 2]:

1. Madelung's quantum hydrodynamics is equivalent to the Schrödinger equation (SE) and leads to description of the quantum particle evolution in the form of Euler equation and continuity equation.
2. SE is consequence of the Liouville equation as result of the local approximation of non-local equations.
3. Generalized Boltzmann physical kinetics leads to the strict approximation of non-local effects in space and time and after transmission to the local approximation leads to parameter $\tau$, which on the quantum level corresponds to the uncertainty principle "time-energy".
4. GHE lead to SE as a deep particular case of the generalized Boltzmann physical kinetics and therefore of non-local hydrodynamics.

In principal CHE needn't in using of the "time-energy" uncertainty relation for estimation of the value of the non-locality parameter $\tau$. Moreover the "time-energy" uncertainty relation does not lead to the exact relations and from position of non-local physics is only the simplest estimation of the non-local effects. Really, let us consider two neighboring physically infinitely small volumes $\mathbf{PhSV_1}$ and $\mathbf{PhSV_2}$ in a nonequilibrium system. Obviously the time $\tau$ should tends to diminish with increasing of the velocities $u$ of particles invading in the nearest neighboring physically infinitely small volume ($\mathbf{PhSV_1}$ or $\mathbf{PhSV_2}$):

$$\tau = H/u^n. \tag{1.7}$$

But the value $\tau$ cannot depend on the velocity direction and naturally to tie $\tau$ with the particle kinetic energy, then

$$\tau = H/{mu^2}, \qquad (1.8)$$

where $H$ is a coefficient of proportionality, which reflects the state of physical system. In the simplest case $H$ is equal to Plank constant $\hbar$ and relation (1.8) became compatible with the Heisenberg relation.

Strict consideration leads to the following system of the generalized hydrodynamic equations (GHE) [4] written in the generalized Euler form:

Continuity equation for species $\alpha$:

$$\frac{\partial}{\partial t}\left\{\rho_\alpha - \tau_\alpha^{(0)}\left[\frac{\partial \rho_\alpha}{\partial t} + \frac{\partial}{\partial \mathbf{r}}\cdot(\rho_\alpha \mathbf{v}_0)\right]\right\} + \frac{\partial}{\partial \mathbf{r}}\cdot\left\{\rho_\alpha \mathbf{v}_0 - \tau_\alpha^{(0)}\left[\frac{\partial}{\partial t}(\rho_\alpha \mathbf{v}_0) + \right.\right.$$
$$\left.\left. + \frac{\partial}{\partial \mathbf{r}}\cdot(\rho_\alpha \mathbf{v}_0 \mathbf{v}_0) + \vec{I}\cdot\frac{\partial p_\alpha}{\partial \mathbf{r}} - \rho_\alpha \mathbf{F}_\alpha^{(1)} - \frac{q_\alpha}{m_\alpha}\rho_\alpha \mathbf{v}_0 \times \mathbf{B}\right]\right\} = R_\alpha, \qquad (1.9)$$

Continuity equation for mixture:

$$\frac{\partial}{\partial t}\left\{\rho - \sum_\alpha \tau_\alpha^{(0)}\left[\frac{\partial \rho_\alpha}{\partial t} + \frac{\partial}{\partial \mathbf{r}}\cdot(\rho_\alpha \mathbf{v}_0)\right]\right\} +$$
$$+ \frac{\partial}{\partial \mathbf{r}}\cdot\left\{\rho\mathbf{v}_0 - \sum_\alpha \tau_\alpha^{(0)}\left[\frac{\partial}{\partial t}(\rho_\alpha \mathbf{v}_0) + \frac{\partial}{\partial \mathbf{r}}\cdot(\rho_\alpha \mathbf{v}_0 \mathbf{v}_0) + \vec{I}\cdot\frac{\partial p_\alpha}{\partial \mathbf{r}} - \right.\right. \qquad (1.10)$$
$$\left.\left. - \rho_\alpha \mathbf{F}_\alpha^{(1)} - \frac{q_\alpha}{m_\alpha}\rho_\alpha \mathbf{v}_0 \times \mathbf{B}\right]\right\} = 0,$$

Momentum equation

$$\frac{\partial}{\partial t}\left\{\rho_\alpha \mathbf{v}_0 - \tau_\alpha^{(0)}\left[\frac{\partial}{\partial t}(\rho_\alpha \mathbf{v}_0) + \frac{\partial}{\partial \mathbf{r}}\cdot \rho_\alpha \mathbf{v}_0 \mathbf{v}_0 + \frac{\partial p_\alpha}{\partial \mathbf{r}} - \rho_\alpha \mathbf{F}_\alpha^{(1)} - \right.\right.$$
$$\left.\left. - \left(\frac{q_\alpha}{m_\alpha}\right)\rho_\alpha \mathbf{v}_0 \times \mathbf{B}\right]\right\} - \mathbf{F}_\alpha^{(1)}\left[\rho_\alpha - \tau_\alpha^{(0)}\left(\frac{\partial \rho_\alpha}{\partial t} + \frac{\partial}{\partial \mathbf{r}}\cdot(\rho_\alpha \mathbf{v}_0)\right)\right] -$$
$$- \frac{q_\alpha}{m_\alpha}\left\{\rho_\alpha \mathbf{v}_0 - \tau_\alpha^{(0)}\left[\frac{\partial}{\partial t}(\rho_\alpha \mathbf{v}_0) + \frac{\partial}{\partial \mathbf{r}}\cdot \rho_\alpha \mathbf{v}_0 \mathbf{v}_0 + \frac{\partial p_\alpha}{\partial \mathbf{r}} - \rho_\alpha \mathbf{F}_\alpha^{(1)} - \right.\right.$$
$$\left.\left. - \frac{q_\alpha}{m_\alpha}\rho_\alpha \mathbf{v}_0 \times \mathbf{B}\right]\right\} \times \mathbf{B} + \frac{\partial}{\partial \mathbf{r}}\cdot\left\{\rho_\alpha \mathbf{v}_0 \mathbf{v}_0 + p_\alpha \vec{I} - \tau_\alpha^{(0)}\left[\frac{\partial}{\partial t}(\rho_\alpha \mathbf{v}_0 \mathbf{v}_0 + \right.\right.\right.$$
$$\left. + p_\alpha \vec{I}) + \frac{\partial}{\partial \mathbf{r}}\cdot\left(\rho_\alpha(\mathbf{v}_0 \mathbf{v}_0)\mathbf{v}_0 + \rho_\alpha(\mathbf{v}_0 \overline{\mathbf{V}_\alpha)\mathbf{V}_\alpha} + \rho_\alpha \overline{(\mathbf{V}_\alpha \mathbf{v}_0)\mathbf{V}_\alpha} + \right.\right.$$
$$\left. + \rho_\alpha \overline{(\mathbf{V}_\alpha \mathbf{V}_\alpha)}\mathbf{v}_0\right) - \mathbf{F}_\alpha^{(1)}\rho_\alpha \mathbf{v}_0 - \rho_\alpha \mathbf{v}_0 \mathbf{F}_\alpha^{(1)} -$$
$$- \frac{q_\alpha}{m_\alpha}\rho_\alpha [\mathbf{v}_0 \times \mathbf{B}]\mathbf{v}_0 - \frac{q_\alpha}{m_\alpha}\rho_\alpha \overline{[\mathbf{V}_\alpha \times \mathbf{B}]\mathbf{V}_\alpha} -$$
$$- \frac{q_\alpha}{m_\alpha}\rho_\alpha \mathbf{v}_0 [\mathbf{v}_0 \times \mathbf{B}] - \frac{q_\alpha}{m_\alpha}\rho_\alpha \overline{\mathbf{V}_\alpha [\mathbf{V}_\alpha \times \mathbf{B}]}\right]\right\} =$$
$$= \int m_\alpha \mathbf{v}_\alpha J_\alpha^{st,el} d\mathbf{v}_\alpha + \int m_\alpha \mathbf{v}_\alpha J_\alpha^{st,inel} d\mathbf{v}_\alpha. \qquad (1.11)$$

Energy equation for mixture:

$$\frac{\partial}{\partial t}\left\{\frac{\rho v_0^2}{2}+\frac{3}{2}p+\sum_\alpha \varepsilon_\alpha n_\alpha - \sum_\alpha \tau_\alpha^{(0)}\left[\frac{\partial}{\partial t}\left(\frac{\rho_\alpha v_0^2}{2}+\frac{3}{2}p_\alpha+\varepsilon_\alpha n_\alpha\right)+\right.\right.$$

$$\left.+\frac{\partial}{\partial \mathbf{r}}\cdot\left(\frac{1}{2}\rho_\alpha v_0^2 \mathbf{v}_0 +\frac{5}{2}p_\alpha \mathbf{v}_0 +\varepsilon_\alpha n_\alpha \mathbf{v}_0\right)-\mathbf{F}_\alpha^{(1)}\cdot\rho_\alpha \mathbf{v}_0\right]\right\}+$$

$$+\frac{\partial}{\partial \mathbf{r}}\cdot\left\{\frac{1}{2}\rho v_0^2 \mathbf{v}_0 +\frac{5}{2}p\mathbf{v}_0 +\mathbf{v}_0\sum_\alpha \varepsilon_\alpha n_\alpha - \sum_\alpha \tau_\alpha^{(0)}\left[\frac{\partial}{\partial t}\left(\frac{1}{2}\rho_\alpha v_0^2 \mathbf{v}_0 +\right.\right.\right.$$

$$\left.+\frac{5}{2}p_\alpha \mathbf{v}_0 +\varepsilon_\alpha n_\alpha \mathbf{v}_0\right)+\frac{\partial}{\partial \mathbf{r}}\cdot\left(\frac{1}{2}\rho_\alpha v_0^2 \mathbf{v}_0 \mathbf{v}_0 +\frac{7}{2}p_\alpha \mathbf{v}_0 \mathbf{v}_0 +\frac{1}{2}p_\alpha v_0^2 \vec{I}+\right.$$

$$\left.+\frac{5}{2}\frac{p_\alpha^2}{\rho_\alpha}\vec{I}+\varepsilon_\alpha n_\alpha \mathbf{v}_0 \mathbf{v}_0 +\varepsilon_\alpha \frac{p_\alpha}{m_\alpha}\vec{I}\right)-\rho_\alpha \mathbf{F}_\alpha^{(1)}\cdot\mathbf{v}_0 \mathbf{v}_0 - p_\alpha \mathbf{F}_\alpha^{(1)}\cdot\vec{I}-$$

$$-\frac{1}{2}\rho_\alpha v_0^2 \mathbf{F}_\alpha^{(1)}-\frac{3}{2}\mathbf{F}_\alpha^{(1)}p_\alpha - \frac{\rho_\alpha v_0^2}{2}\frac{q_\alpha}{m_\alpha}[\mathbf{v}_0\times\mathbf{B}]-\frac{5}{2}p_\alpha\frac{q_\alpha}{m_\alpha}[\mathbf{v}_0\times\mathbf{B}]-$$

$$\left.\left.-\varepsilon_\alpha n_\alpha \frac{q_\alpha}{m_\alpha}[\mathbf{v}_0\times\mathbf{B}]-\varepsilon_\alpha n_\alpha \mathbf{F}_\alpha^{(1)}\right]\right\}-\left\{\mathbf{v}_0\cdot\sum_\alpha \rho_\alpha \mathbf{F}_\alpha^{(1)}-\sum_\alpha \tau_\alpha^{(0)}\left[\mathbf{F}_\alpha^{(1)}\cdot\right.\right.$$

$$\left.\left.\cdot\left(\frac{\partial}{\partial t}(\rho_\alpha \mathbf{v}_0)+\frac{\partial}{\partial \mathbf{r}}\cdot\rho_\alpha \mathbf{v}_0 \mathbf{v}_0 +\frac{\partial}{\partial \mathbf{r}}\cdot p_\alpha \vec{I}-\rho_\alpha \mathbf{F}_\alpha^{(1)}-q_\alpha n_\alpha[\mathbf{v}_0\times\mathbf{B}]\right)\right]\right\}=0. \qquad (1.12)$$

Here $\mathbf{F}_\alpha^{(1)}$ are the forces of the non-magnetic origin, $\mathbf{B}$ - magnetic induction, $\vec{I}$ - unit tensor, $q_\alpha$ - charge of the $\alpha$-component particle, $p_\alpha$ - static pressure for $\alpha$-component, $\mathbf{V}_\alpha$ - thermal velocity, $\varepsilon_\alpha$ - internal energy for the particles of $\alpha$-component, $\mathbf{v}_0$ - hydrodynamic velocity for mixture.

## 2. Nonstationary 1D generalized hydrodynamic equations in the self consistent electrical field. Quantization in the generalized quantum hydrodynamics.

In the following we intend to obtain the soliton's type of solution of the generalized hydrodynamic equations for plasma in the self consistent electrical field. All elements of possible formation like quantum soliton should move with the same translational velocity. Then the system of GHE consist from the generalized Poisson equation reflecting the effects of the charge and the charge flux perturbations, two continuity equations for positive and negative species (in particular, for ion and electron components), one motion equation and two energy equations for ion and electron components. This system of six equations for non-stationary 1D case can be written in the form [1 - 4]:
(Poisson equation)

$$\frac{\partial^2 \varphi}{\partial x^2}=-4\pi e\left\{\left[n_i - \tau_i\left(\frac{\partial n_i}{\partial t}+\frac{\partial}{\partial x}(n_i u)\right)\right]-\left[n_e - \tau_e\left(\frac{\partial n_e}{\partial t}+\frac{\partial}{\partial x}(n_e u)\right)\right]\right\}, \qquad (2.1)$$

(continuity equation for positive ion component)

$$\frac{\partial}{\partial t}\left\{\rho_i - \tau_i\left[\frac{\partial \rho_i}{\partial t}+\frac{\partial}{\partial x}(\rho_i u)\right]\right\}+\frac{\partial}{\partial x}\left\{\rho_i u - \tau_i\left[\frac{\partial}{\partial t}(\rho_i u)+\frac{\partial}{\partial x}(\rho_i u^2)+\frac{\partial p_i}{\partial x}-\rho_i F_i\right]\right\}, \qquad (2.2)$$

(continuity equation for electron component)

$$\frac{\partial}{\partial t}\left\{\rho_e - \tau_e\left[\frac{\partial \rho_e}{\partial t} + \frac{\partial}{\partial x}(\rho_e u)\right]\right\} + \frac{\partial}{\partial x}\left\{\rho_e u - \tau_e\left[\frac{\partial}{\partial t}(\rho_e u) + \frac{\partial}{\partial x}(\rho_e u^2) + \frac{\partial p_e}{\partial x} - \rho_e F_e\right]\right\}, \qquad (2.3)$$

(momentum equation)

$$\frac{\partial}{\partial t}\left\{\rho u - \tau_i\left[\frac{\partial}{\partial t}(\rho_i u) + \frac{\partial}{\partial x}(p_i + \rho_i u^2) - \rho_i F_i\right] - \tau_e\left[\frac{\partial}{\partial t}(\rho_e u) + \frac{\partial}{\partial x}(p_e + \rho_e u^2) - \rho_e F_e\right]\right\} -$$

$$- \rho_i F_i - \rho_e F_e + F_i \tau_i \left(\frac{\partial \rho_i}{\partial t} + \frac{\partial}{\partial x}(\rho_i u)\right) + F_e \tau_e \left(\frac{\partial \rho_e}{\partial t} + \frac{\partial}{\partial x}(\rho_e u)\right) + \qquad (2.4)$$

$$+ \frac{\partial}{\partial x}\left\{\begin{array}{l}\rho u^2 + p - \tau_i\left[\frac{\partial}{\partial t}(\rho_i u^2 + p_i) + \frac{\partial}{\partial x}(\rho_i u^3 + 3p_i u) - 2\rho_i u F_i\right] - \\ - \tau_e\left[\frac{\partial}{\partial t}(\rho_e u^2 + p_e) + \frac{\partial}{\partial x}(\rho_e u^3 + 3p_e u)\right] - 2\rho_e u F_e\end{array}\right\} = 0$$

(energy equation for positive ion component)

$$\frac{\partial}{\partial t}\left\{\rho_i u^2 + 3p_i - \tau_i\left[\frac{\partial}{\partial t}(\rho_i u^2 + 3p_i) + \frac{\partial}{\partial x}(\rho_i u^3 + 5p_i u) - 2\rho_i F_i u\right]\right\} +$$

$$+ \frac{\partial}{\partial x}\left\{\rho_i u^3 + 5p_i u - \tau_i\left[\frac{\partial}{\partial t}(\rho_i u^3 + 5p_i u) + \frac{\partial}{\partial x}\left(\rho_i u^4 + 8p_i u^2 + 5\frac{p_i^2}{\rho_i}\right) - F_i(3\rho_i u^2 + 5p_i)\right]\right\}, (2.5)$$

$$- 2u\rho_i F_i + 2\tau_i F_i\left[\frac{\partial}{\partial t}(\rho_i u) + \frac{\partial}{\partial x}(\rho_i u^2 + p_i) - \rho_i F_i\right] = -\frac{p_i - p_e}{\tau_{ei}}$$

(energy equation for electron component)

$$\frac{\partial}{\partial t}\left\{\rho_e u^2 + 3p_e - \tau_e\left[\frac{\partial}{\partial t}(\rho_e u^2 + 3p_e) + \frac{\partial}{\partial x}(\rho_e u^3 + 5p_e u) - 2\rho_e F_e u\right]\right\} +$$

$$+ \frac{\partial}{\partial x}\left\{\rho_e u^3 + 5p_e u - \tau_e\left[\frac{\partial}{\partial t}(\rho_e u^3 + 5p_e u) + \frac{\partial}{\partial x}\left(\rho_e u^4 + 8p_e u^2 + 5\frac{p_e^2}{\rho_e}\right) - F_e(3\rho_e u^2 + 5p_e)\right]\right\},(2.6)$$

$$- 2u\rho_e F_e + 2\tau_e F_e\left[\frac{\partial}{\partial t}(\rho_e u) + \frac{\partial}{\partial x}(\rho_e u^2 + p_e) - \rho_e F_e\right] = -\frac{p_e - p_i}{\tau_{ei}}$$

where $u$ is translational velocity of the quantum object, $\varphi$ - scalar potential, $n_i$ and $n_e$ are the number density of the charged species, $F_i$ and $F_e$ are the forces acting on the unit mass of ion and electron.

Approximations for non-local parameters $\tau_i$, $\tau_e$ and $\tau_{ei}$ need the special consideration. In the following for the $\tau_i$ and $\tau_i$ approximation the relation (1.8) is used in the forms

$$\tau_i = H\!\left/m_i u^2\right., \quad \tau_e = H\!\left/m_e u^2\right.. \tag{2.7}$$

For non-local parameter of electron-ion interaction $\tau_{ei}$ is applicable the relation

$$\frac{1}{\tau_{ei}} = \frac{1}{\tau_e} + \frac{1}{\tau_i}. \tag{2.8}$$

In this case parameter $\tau_{ei}$ serves as relaxation time in the process of the particle interaction of different kinds. Transformation (2.8) for the case $H = \hbar$ leads to the obvious compatibility with the Heisenberg principle

$$\frac{1}{\tau_{ei}} = \frac{\tau_e + \tau_i}{\tau_e \tau_i} = \frac{\dfrac{\hbar}{m_e u^2} + \dfrac{\hbar}{m_i u^2}}{\dfrac{\hbar^2}{u^4}\dfrac{1}{m_e m_i}} = \frac{u^2}{\hbar}(m_e + m_i). \tag{2.9}$$

Then

$$u^2 (m_e + m_i)\tau_{ei} = \hbar. \tag{2.10}$$

Equality (2.10) is consequence of "time-energy" uncertainty relation for combined particle with mass $m_i + m_e$.

In principal the time values $\tau_i$ and $\tau_e$ should be considered as sums of mean times between collisions ($\tau_i^{tr}, \tau_e^{tr}$) and discussed above non-local quantum values ($\tau_i^{qu}, \tau_e^{qu}$), namely for example

$$\tau_i = \tau_i^{tr} + \tau_i^{qu}. \tag{2.11}$$

For molecular hydrogen in standard conditions mean time between collisions is equal to $6.6 \cdot 10^{-11}$ s. For quantum objects moving with velocities typical for lightning balls $\tau^{qu}$ is much more than $\tau^{tr}$ and the usual static pressure $p$ transforms in the pressure which can be named as the rest non-local pressure. In the definite sense this kind of pressure can be considered as analogue of the Bose condensate pressure.

The following formulae are valid for acting forces

$$F_i = -\frac{e}{m_i}\frac{\partial \varphi}{\partial x}, \quad F_e = -\frac{e}{m_e}\frac{\partial \varphi}{\partial x}. \tag{2.12}$$

Let us consider now the introduction of quantization in quantum hydrodynamics. With this aim write down the expression for the total energy $E$ of point-like particle moving along the positive direction of $x$-axis with velocity $u$ in the attractive field of Coulomb forces

$$E = \frac{mu^2}{2} - \frac{Ze^2}{x}, \tag{2.13}$$

where $Z$ is the charge number and $x$ is the distance from the center of forces. If this movement obeys to the condition of non-locality $mu^2 = H/\tau$ and $x = u\tau$, then

$$E = \frac{H^2}{2mx^2} - \frac{Ze^2}{x}. \tag{2.14}$$

Minimal total energy corresponds to the condition $\left(\frac{\partial E}{\partial x}\right)_{x=x_B} = 0$ and

$$\frac{H^2}{mx_B^3} = \frac{Ze^2}{x_B^2}. \tag{2.15}$$

From (2.15) follows

$$x_B = \frac{H^2}{Zme^2}. \tag{2.16}$$

and from (2.14), (2.16)

$$E = \frac{H^2}{2m}\frac{Z^2m^2e^4}{H^4} - Ze^2\frac{Zme^2}{H^2} = -\frac{Z^2me^4}{2H^2}. \tag{2.17}$$

For atom with single electron in the shell moving on the Bohr's orbit of radius $r_B$, Eq. (2.15) with taking into account the relation

$$\frac{H^2}{mr_B^3} = \frac{mu^2}{r_B^2} \tag{2.18}$$

leads to equality of Coulomb and inertial forces for the orbit electron

$$\frac{m_e u^2}{r_B} = \frac{Ze^2}{r_B^2} \tag{2.19}$$

and to the same expression for energy

$$E = -\frac{Z^2 m_e e^4}{2H^2}. \tag{2.20}$$

The comparison of Eq. (2.20) with the Balmer's relation leads to condition

$$H = n\hbar \tag{2.21}$$

with integer $n = 1, 2, ...$ known as principal quantum number and well known relation

$$E = -\frac{Z^2 m_e e^4}{2\hbar^2}\frac{1}{n^2}. \tag{2.22}$$

Eqs. (2.16), (2.19) lead to the character velocity for this obviously model problem

$$u = \frac{Ze^2}{H} \tag{2.23}$$

with the velocity $2.187 \cdot 10^8$ cm/s.

As we see the mentioned simple considerations allow in principal to introduce quantization in the quantum hydrodynamics without direct application of Schrödinger equation. Important to notice that conditions of quantization are not the intrinsic feature of Schrödinger equation, for example the appearance of quantization in Schrödinger's theory is connected with the truncation of infinite series and transformation in polynomials with the finite quantity of terms.

### 3. Quantum solitons in self consistent electric field.

Let us introduce the coordinate system moving along the positive direction of $x$- axis in ID space with velocity $C = u_0$ equal to phase velocity of considering quantum object

$$\xi = x - Ct. \qquad (3.1)$$

Taking into account the De Broglie relation we should wait that the group velocity $u_g$ is equal $2u_0$. In moving coordinate system all dependent hydrodynamic values are function of $(\xi, t)$. We investigate the possibility of the quantum object formation of the soliton type. For this solution there is no explicit dependence on time for coordinate system moving with the phase velocity $u_0$. Write down the system of equations (2.1) - (2.6) for the two component mixture of charged particles without taking into account the component's internal energy in the dimensionless form, where dimensionless symbols are marked by tildes. We begin with introduction the scales for velocity

$$[u] = u_0 \qquad (3.2)$$

and for coordinate $x$

$$\frac{\hbar}{m_e u_0} = x_0. \qquad (3.3)$$

Generalized Poisson equation (2.1)

$$\frac{\partial^2 \varphi}{\partial x^2} = -4\pi e \left\{ \left[ n_i - \frac{\hbar}{m_i u^2} u_0 \left( -\frac{\partial n_i}{\partial x} + \frac{\partial}{\partial x}(n_i \tilde{u}) \right) \right] - \left[ n_e - \frac{\hbar}{m_e u^2} u_0 \left( -\frac{\partial n_e}{\partial x} + \frac{\partial}{\partial x}(n_e \tilde{u}) \right) \right] \right\} \qquad (3.4)$$

now is written as

$$\frac{\partial^2 \tilde{\varphi}}{\partial \tilde{\xi}^2} = -\left\{ \frac{m_e}{m_i} \left[ \tilde{\rho}_i - \frac{1}{\tilde{u}^2} \frac{m_e}{m_i} \left( -\frac{\partial \tilde{\rho}_i}{\partial \tilde{\xi}} + \frac{\partial}{\partial \tilde{\xi}}(\tilde{\rho}_i \tilde{u}) \right) \right] - \left[ \tilde{\rho}_e - \frac{1}{\tilde{u}^2} \left( -\frac{\partial \tilde{\rho}_e}{\partial \tilde{\xi}} + \frac{\partial}{\partial \tilde{\xi}}(\tilde{\rho}_e \tilde{u}) \right) \right] \right\},$$

if the potential scale $\varphi_0$ and the density scale $\rho_0$ are chosen as

$$\varphi_0 = \frac{m_e}{e} u_0^2, \qquad (3.5)$$

$$\rho_0 = \frac{m_e^4}{4\pi \hbar^2 e^2} u_0^4. \qquad (3.6)$$

Scaled forces will be described by ($e$ - absolute electron charge) relations

$$\rho_i F_i = -\frac{u_0^2}{x_0} \rho_0 \frac{m_e}{m_i} \frac{\partial \tilde{\varphi}}{\partial \tilde{\xi}} \tilde{\rho}_i, \qquad (3.7)$$

$$\rho_e F_e = \frac{u_0^2}{x_0} \rho_0 \frac{\partial \tilde{\varphi}}{\partial \tilde{\xi}} \tilde{\rho}_e. \qquad (3.8)$$

Analogical transformations should be applied to the other equations of the system (2.1) - (2.6). We have the following system of six ordinary non-linear equations

$$\frac{\partial^2 \tilde{\varphi}}{\partial \tilde{\xi}^2} = -\left\{ \frac{m_e}{m_i}\left[\tilde{\rho}_i - \frac{1}{\tilde{u}^2}\frac{m_e}{m_i}\left(-\frac{\partial \tilde{\rho}_i}{\partial \tilde{\xi}} + \frac{\partial}{\partial \tilde{\xi}}(\tilde{\rho}_i \tilde{u})\right)\right] - \left[\tilde{\rho}_e - \frac{1}{\tilde{u}^2}\left(-\frac{\partial \tilde{\rho}_e}{\partial \tilde{\xi}} + \frac{\partial}{\partial \tilde{\xi}}(\tilde{\rho}_e \tilde{u})\right)\right]\right\}, \qquad (3.9)$$

$$\frac{\partial \tilde{\rho}_i}{\partial \tilde{\xi}} - \frac{\partial \tilde{\rho}_i \tilde{u}}{\partial \tilde{\xi}} + \frac{m_e}{m_i}\frac{\partial}{\partial \tilde{\xi}}\left\{\frac{1}{\tilde{u}^2}\left[\frac{\partial}{\partial \tilde{\xi}}(\tilde{p}_i + \tilde{\rho}_i + \tilde{\rho}_i \tilde{u}^2 - 2\tilde{\rho}_i \tilde{u}_i) + \frac{m_e}{m_i}\tilde{\rho}_i \frac{\partial \tilde{\varphi}}{\partial \tilde{\xi}}\right]\right\} = 0, \quad (3.10)$$

$$\frac{\partial \tilde{\rho}_e}{\partial \tilde{\xi}} - \frac{\partial \tilde{\rho}_e \tilde{u}}{\partial \tilde{\xi}} + \frac{\partial}{\partial \tilde{\xi}}\left\{\frac{1}{\tilde{u}^2}\left[\frac{\partial}{\partial \tilde{\xi}}(\tilde{p}_e + \tilde{\rho}_e + \tilde{\rho}_e \tilde{u}^2 - 2\tilde{\rho}_e \tilde{u}_e) - \tilde{\rho}_e \frac{\partial \tilde{\varphi}}{\partial \tilde{\xi}}\right]\right\} = 0, \qquad (3.11)$$

$$\frac{\partial}{\partial \tilde{\xi}}\left\{(\tilde{\rho}_i + \tilde{\rho}_e)\tilde{u}^2 + (\tilde{p}_i + \tilde{p}_e) - (\tilde{\rho}_i + \tilde{\rho}_e)\tilde{u}\right\} +$$

$$+ \frac{\partial}{\partial \tilde{\xi}}\left\{ \begin{aligned} &\frac{1}{\tilde{u}^2}\frac{m_e}{m_i}\left[\frac{\partial}{\partial \tilde{\xi}}(2\tilde{p}_i + 2\tilde{\rho}_i \tilde{u}^2 - \tilde{\rho}_i \tilde{u} - \tilde{\rho}_i \tilde{u}^3 - 3\tilde{p}_i \tilde{u}) + \tilde{\rho}_i \frac{m_e}{m_i}\frac{\partial \tilde{\varphi}}{\partial \tilde{\xi}}\right] + \\ &+ \frac{1}{\tilde{u}^2}\left[\frac{\partial}{\partial \tilde{\xi}}(2\tilde{p}_e + 2\tilde{\rho}_e \tilde{u}^2 - \tilde{\rho}_e \tilde{u} - \tilde{\rho}_e \tilde{u}^3 - 3\tilde{p}_e \tilde{u}) - \tilde{\rho}_e \frac{\partial \tilde{\varphi}}{\partial \tilde{\xi}}\right] \end{aligned}\right\} +$$

$$+ \tilde{\rho}_i \frac{m_e}{m_i}\frac{\partial \tilde{\varphi}}{\partial \tilde{\xi}} - \tilde{\rho}_e \frac{\partial \tilde{\varphi}}{\partial \tilde{\xi}} - \frac{\partial \tilde{\varphi}}{\partial \tilde{\xi}}\frac{1}{\tilde{u}^2}\left(\frac{m_e}{m_i}\right)^2\left(-\frac{\partial \tilde{\rho}_i}{\partial \tilde{\xi}} + \frac{\partial}{\partial \tilde{\xi}}(\tilde{\rho}_i \tilde{u})\right) +$$

$$+ \frac{\partial \tilde{\varphi}}{\partial \tilde{\xi}}\frac{1}{\tilde{u}^2}\left(-\frac{\partial \tilde{\rho}_e}{\partial \tilde{\xi}} + \frac{\partial}{\partial \tilde{\xi}}(\tilde{\rho}_e \tilde{u})\right) - 2\frac{\partial}{\partial \tilde{\xi}}\left\{\frac{1}{\tilde{u}}\frac{\partial \tilde{\varphi}}{\partial \tilde{\xi}}\left[\left(\frac{m_e}{m_i}\right)^2 \tilde{\rho}_i - \tilde{\rho}_e\right]\right\} = 0, \qquad (3.12)$$

$$\frac{\partial}{\partial \tilde{\xi}}\left\{\tilde{\rho}_i \tilde{u}^3 + 5\tilde{p}_i \tilde{u} - \tilde{\rho}_i \tilde{u}^2 - 3\tilde{p}_i\right\} +$$

$$+ \frac{\partial}{\partial \tilde{\xi}}\left\{\frac{1}{\tilde{u}^2}\frac{m_e}{m_i}\left[\begin{aligned}&\frac{\partial}{\partial \tilde{\xi}}\left(2\tilde{\rho}_i \tilde{u}^3 + 10\tilde{p}_i \tilde{u} - \tilde{\rho}_i \tilde{u}^4 - 8\tilde{p}_i \tilde{u}^2 - 5\frac{\tilde{p}_i^2}{\tilde{\rho}_i} - \tilde{\rho}_i \tilde{u}^2 - 3\tilde{p}_i\right) + \\ &+ \frac{m_e}{m_i}\frac{\partial \tilde{\varphi}}{\partial \tilde{\xi}}\left(2\tilde{\rho}_i \tilde{u} - 3\tilde{\rho}_i \tilde{u}^2 - 5\tilde{p}_i\right)\end{aligned}\right]\right\}$$

$$+ 2\frac{m_e}{m_i}\tilde{\rho}_i \tilde{u}\frac{\partial \tilde{\varphi}}{\partial \tilde{\xi}} -$$

$$- 2\frac{\partial \tilde{\varphi}}{\partial \tilde{\xi}}\frac{1}{\tilde{u}^2}\left(\frac{m_e}{m_i}\right)^2\left[\frac{\partial}{\partial \tilde{\xi}}(\tilde{\rho}_i \tilde{u}^2 + \tilde{p}_i - \tilde{\rho}_i \tilde{u}) + \tilde{\rho}_i \frac{m_e}{m_i}\frac{\partial \tilde{\varphi}}{\partial \tilde{\xi}}\right] = -(\tilde{p}_i - \tilde{p}_e)\tilde{u}^2\left(1 + \frac{m_i}{m_e}\right). \qquad (3.13)$$

$$\frac{\partial}{\partial \tilde{\xi}}\left\{\tilde{\rho}_e \tilde{u}^3 + 5\tilde{p}_e \tilde{u} - \tilde{\rho}_e \tilde{u}^2 - 3\tilde{p}_e\right\} +$$

$$+ \frac{\partial}{\partial \tilde{\xi}}\left\{\frac{1}{\tilde{u}^2}\left[\begin{array}{l}\frac{\partial}{\partial \tilde{\xi}}\left(2\tilde{\rho}_e \tilde{u}^3 + 10\tilde{p}_e \tilde{u} - \tilde{\rho}_e \tilde{u}^4 - 8\tilde{p}_e \tilde{u}^2 - 5\frac{\tilde{p}_e^2}{\tilde{\rho}_e} - \tilde{\rho}_e \tilde{u}^2 - 3\tilde{p}_e\right) + \\ + \frac{\partial \tilde{\varphi}}{\partial \tilde{\xi}}\left(3\tilde{\rho}_e \tilde{u}^2 + 5\tilde{p}_e - 2\tilde{\rho}_e \tilde{u}\right)\end{array}\right]\right\} -$$

$$-2\tilde{\rho}_e \tilde{u}\frac{\partial \tilde{\varphi}}{\partial \tilde{\xi}} +$$

$$+2\frac{\partial \tilde{\varphi}}{\partial \tilde{\xi}}\frac{1}{\tilde{u}^2}\left[\frac{\partial}{\partial \tilde{\xi}}\left(\tilde{\rho}_e \tilde{u}^2 + \tilde{p}_e - \tilde{\rho}_e \tilde{u}\right) - \tilde{\rho}_e \frac{\partial \tilde{\varphi}}{\partial \tilde{\xi}}\right] = -\left(\tilde{p}_e - \tilde{p}_i\right)\left(1 + \frac{m_i}{m_e}\right)\tilde{u}^2.$$

(3.14)

Some comments to Eqs. (3.9) – (3.14):
1. Every equation from the system is of the second order and needs two conditions. The problem belongs to the class of Cauchy problems.
2. In comparison with the Schrödinger theory connected with behavior of the wave function, no special conditions are applied for dependent variables including the domain of the solution existing. This domain is defined automatically in the process of the numerical solution of the concrete variant of calculations.
3. From the introduced scales

$$u_0, \; x_0 = \frac{\hbar}{m_e}\frac{1}{u_0}, \; \varphi_0 = \frac{m_e}{e}u_0^2, \; \rho_0 = \frac{m_e^4}{4\pi\hbar^2 e^2}u_0^4, \; p_0 = \rho_0 u_0^2 = \frac{m_e^4}{4\pi\hbar^2 e^2}u_0^6$$

only two parameters are independent – the phase velocity $u_0$ of the quantum object, and external parameter $H$, which is proportional to Plank constant $\hbar$ and in general case should be inserted in the scale relation as $x_0 = \frac{H}{m_e u_0} = \frac{n\hbar}{m_e u_0}$. It leads to exchange in all scales $\hbar \leftrightarrow H$. But the value $v^{qu} = \hbar/m_e$ has the dimension $[cm^2/s]$ and can be titled as quantum viscosity, $v^{qu} = 1.1577 \; cm^2/s$. Of course in principal the electron mass can be replaced in scales by mass of other particles with the negative charge. From this point of view the obtained solutions which will be discussed below have the universal character defined only by Cauchy conditions.

**4. Results of mathematical modeling.**

The system of generalized quantum hydrodynamic equations (3.9) – (3.14) have the great possibilities of mathematical modeling as result of changing of twelve Cauchy conditions describing the character features of initial perturbations which lead to the soliton formation.

On this step of investigation I intend to demonstrate the influence of difference conditions on the soliton formation. The following figures reflect some results of calculations realized according to the system of equations (3.9) - (3.14) with the help of Maple 9. The following notations on figures are used: r- density $\tilde{\rho}_i$ (solid black line), s- density $\tilde{\rho}_e$ (solid line), u- velocity $\tilde{u}$ (dashed line), p - pressure $\tilde{p}_i$ (black dash dotted line), q - pressure $\tilde{p}_e$ (dash dotted line) and v - self consistent potential $\tilde{\varphi}$. Explanations placed under all following figures, Maple program contains Maple's notations – for example the expression $D(u)(0) = 0$ means in usual notations $\frac{\partial \tilde{u}}{\partial \tilde{\xi}}(0) = 0$, independent variable $t$ responds to $\tilde{\xi}$.

We begin with investigation of the problem of principle significance – is it possible after a perturbation (defined by Cauchy conditions) to obtain the quantum object of the soliton's kind as result of the self-organization of ionized matter? In the case of the positive answer, what is the origin of existence of this stable object? By the way the mentioned questions belong to the typical problem in the theory of the ball lightning. With this aim let us consider the initial perturbations

```
v(0)=1,r(0)=1,s(0)=1,u(0)=1,p(0)=1,q(0)=.95,
D(v)(0)=0,D(r)(0)=0,D(s)(0)=0,D(u)(0)=0,D(p)(0)=0,D(q)(0)=0
```

in the mixture of positive and negative ions of equal masses if the pressure $\tilde{p}_i(0)$ of positive particles is larger than $\tilde{p}_e(0)$ of the negative ones (for the variant under consideration `p(0)=1,q(0)=.95`). The following figures 1 – 3 reflect the result of solution of Eqs. (3.9) – (3.14).

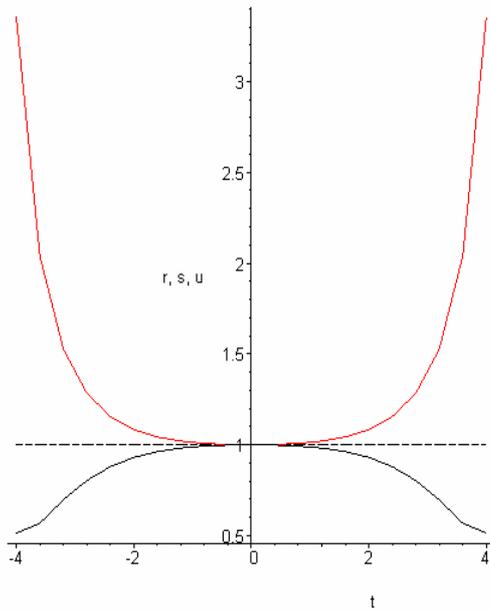 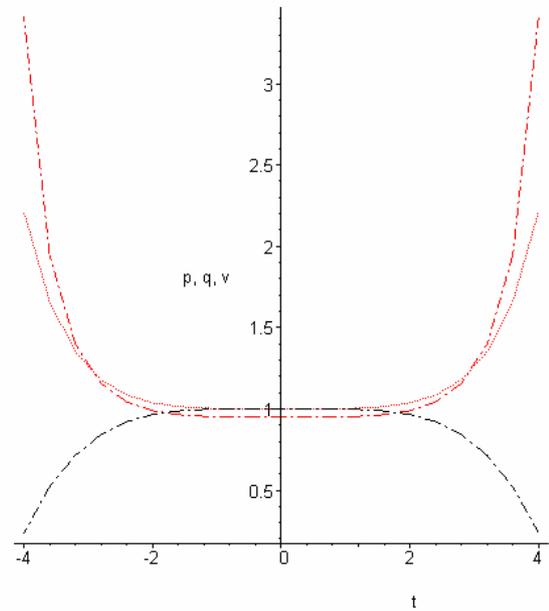

Fig.1. r- density $\tilde{\rho}_i$, u- velocity $\tilde{u}$, s- density $\tilde{\rho}_e$ in quantum soliton.

Fig.2. p - pressure $\tilde{p}_i$, q - pressure $\tilde{p}_e$, v - self consistent potential $\tilde{\varphi}$ in quantum soliton.

Fig. 1 displays the quantum object placed in bounded region of 1D space, all parts of this object are moving with the same velocity. Important to underline that no special boundary conditions were used for this and all following cases. Then this soliton is product of the self-organization of ionized matter. Fig. 3 contains the answer for formulated above question about stability of the object. Really the object is restricted by negative shell. The derivative $\frac{\partial \tilde{\varphi}}{\partial \tilde{\xi}}$ is proportional to the self-consistent forces acting on the positive and negative parts of the soliton. Consider for example the right side of soliton. The self consistent force compresses the positive part of this soliton and provokes the movement of the negative part along the positive direction of the $\tilde{\xi}$ - axis (t – axis in nomination of Fig. 1). But the increasing of quantum pressure prevent to destruction of soliton. Therefore the stability of the quantum object is result of the self-consistent influence of electric potential and quantum pressures.

Interesting to notice that stability can be achieved if soliton has positive shell and negative kernel but $\tilde{p}_i(0) < \tilde{p}_e(0)$, see Fig. 4 – 6 obtained as result of mathematical modeling for the case

```
v(0)=1,r(0)=1,s(0)=1,u(0)=1,p(0)=1,q(0)=1.05,
D(v)(0)=0,D(r)(0)=0,D(s)(0)=0,D(u)(0)=0,D(p)(0)=0,D(q)(0)=0.
```

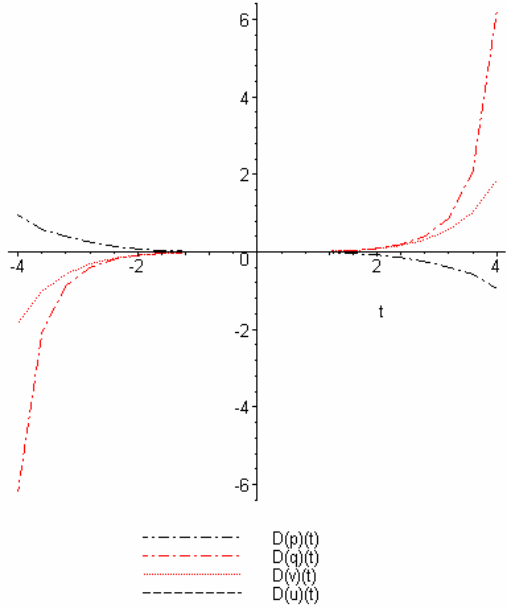

Fig. 3. The derivative of pressure of the positive component $\frac{\partial \tilde{p}_i}{\partial \tilde{\xi}}$, the derivative of pressure of negative component $\frac{\partial \tilde{p}_e}{\partial \tilde{\xi}}$, the derivative of the self-consistent potential $\frac{\partial \tilde{\varphi}}{\partial \tilde{\xi}}$ in quantum soliton.

The explanation for this case has practically the same character as in the previous case but positive and negative species change their roles.

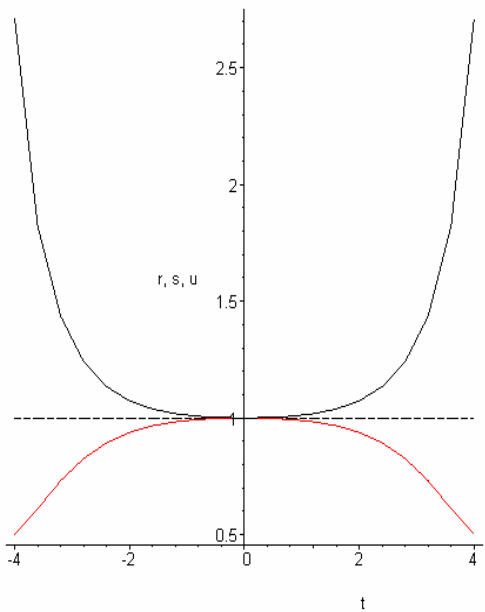 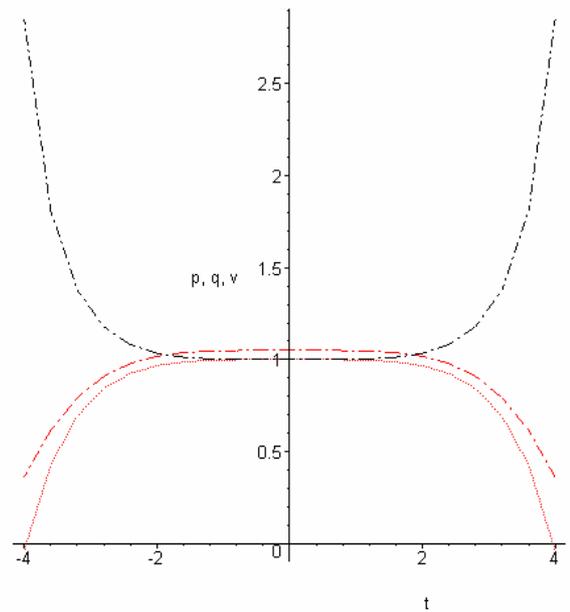

Fig.4. r- density $\tilde{\rho}_i$, u- velocity $\tilde{u}$, s- density $\tilde{\rho}_e$ in quantum soliton.

Fig.5. p - pressure $\tilde{p}_i$, q - pressure $\tilde{p}_e$, v - self consistent potential $\tilde{\varphi}$ in quantum soliton.

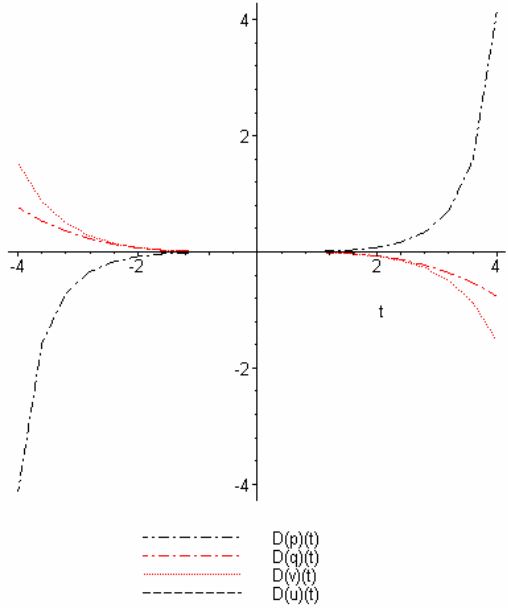

Fig. 6. The derivative of pressure of the positive component $\frac{\partial \tilde{p}_i}{\partial \tilde{\xi}}$, the derivative of pressure of negative component $\frac{\partial \tilde{p}_e}{\partial \tilde{\xi}}$, the derivatives of the self-consistent potential $\frac{\partial \tilde{\varphi}}{\partial \tilde{\xi}}$ and velocity $\frac{\partial \tilde{u}}{\partial \tilde{\xi}}$ in quantum soliton.

Everywhere in following calculations we use the typical ratio of masses $m_i/m_e = 1838$. The initial perturbations in the mixture of heavy positive particles and electrons produce the soliton formation if the pressure $\tilde{p}_i(0)$ of the positive particles is larger than $\tilde{p}_e(0)$ of the negative ones (for the variant under consideration `p(0)=1,q(0)=.95`):
`v(0)=1,r(0)=1,s(0)=1/1838,u(0)=1,p(0)=1,q(0)=.95,`
`D(v)(0)=0,D(r)(0)=0,D(s)(0)=0,D(u)(0)=0,D(p)(0)=0,D(q)(0)=0.`

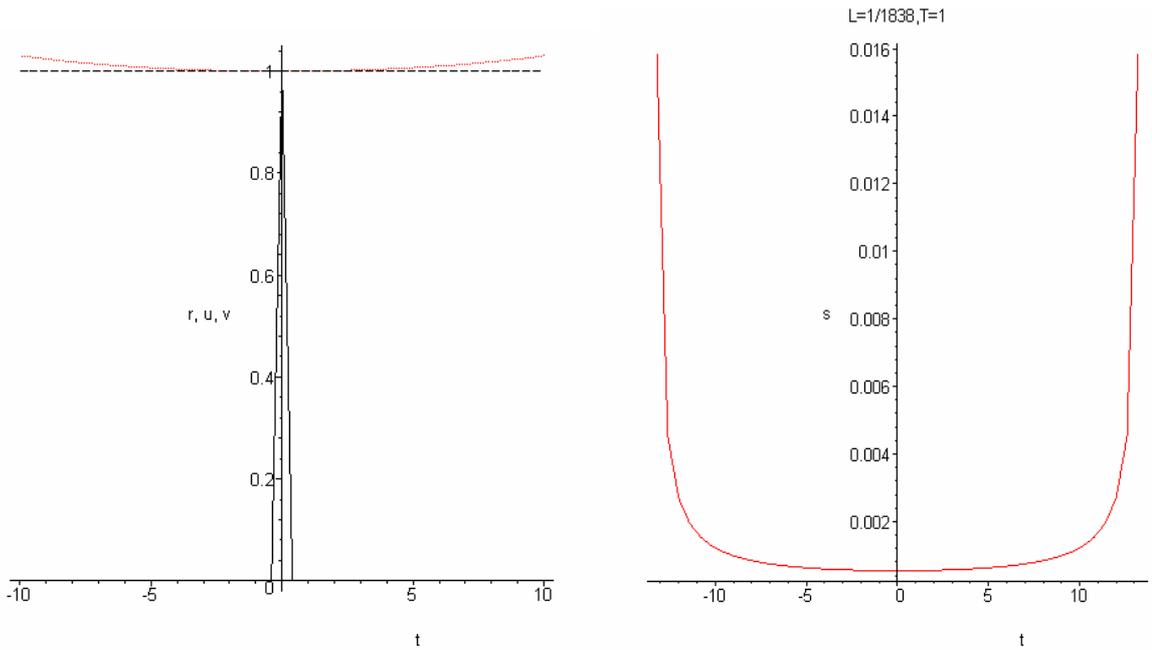

Fig.7. r- density $\tilde{\rho}_i$, u- velocity $\tilde{u}$, v-self-consistent potential $\tilde{\varphi}$ in quantum soliton.  Fig.8. s- density $\tilde{\rho}_e$ in quantum soliton.

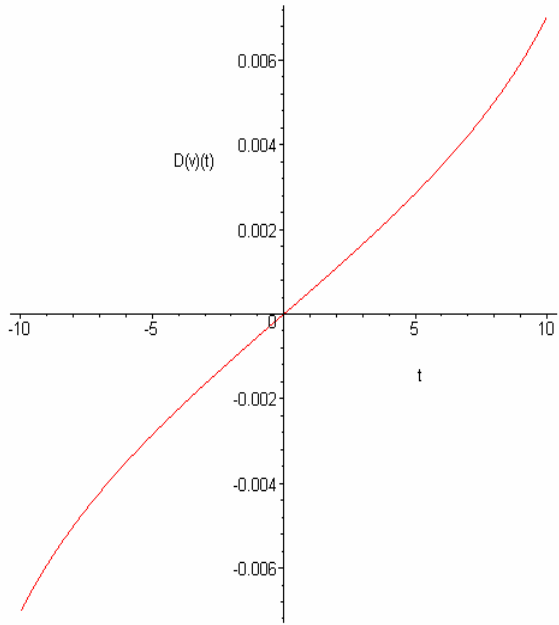
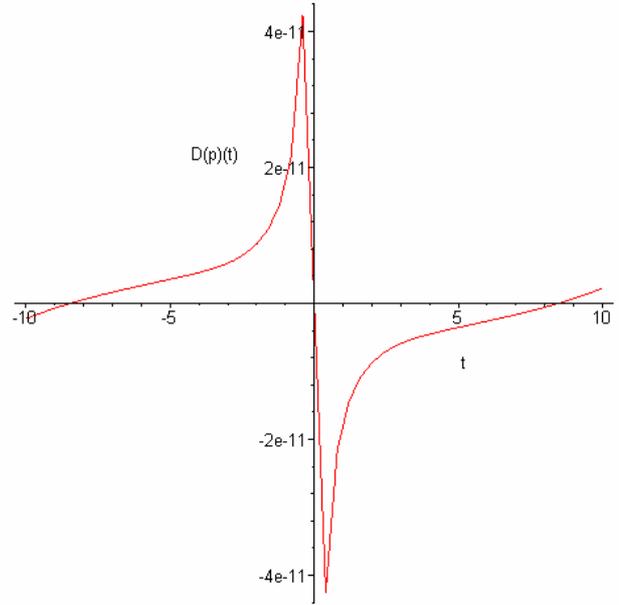

Fig. 9. The derivative of the self-consistent potential $\dfrac{\partial \tilde{\varphi}}{\partial \tilde{\xi}}$ in quantum soliton.

Fig. 10. The derivative of the positive component pressure $\dfrac{\partial \tilde{p}_i}{\partial \tilde{\xi}}$ in quantum soliton.

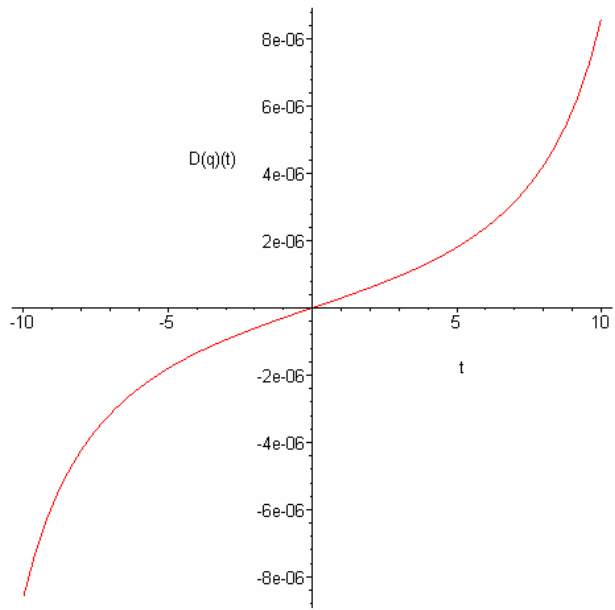

Fig. 11. The derivative of the negative component pressure $\dfrac{\partial \tilde{p}_e}{\partial \tilde{\xi}}$ in quantum soliton.

In comparison with Fig. 1 – 3 we observe the explicit positive kernel which is typical for atom structures.

Now can be demonstrated the influence of the significant difference in mass of particles for the case $\tilde{p}_i(0) < \tilde{p}_e(0)$, the Cauchy conditions

```
v(0)=1,r(0)=1,s(0)=1/1838,u(0)=1,p(0)=1,q(0)=1.05,
D(v)(0)=0,D(r)(0)=0,D(s)(0)=0,D(u)(0)=0,D(p)(0)=0,D(q)(0)=0
```

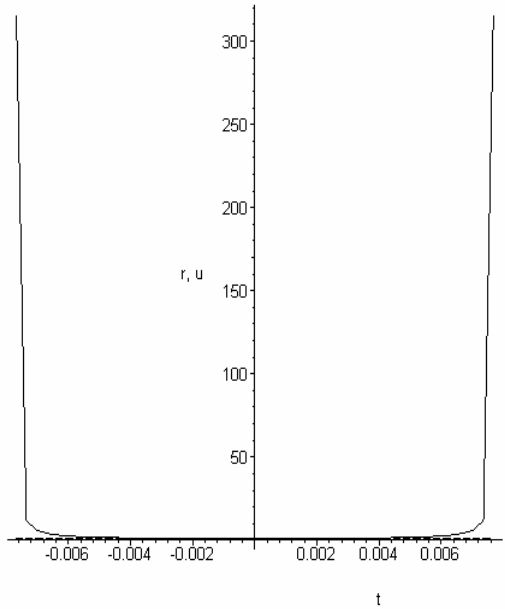

Fig.12. r- density $\tilde{\rho}_i$, u- velocity $\tilde{u}$, in quantum soliton.

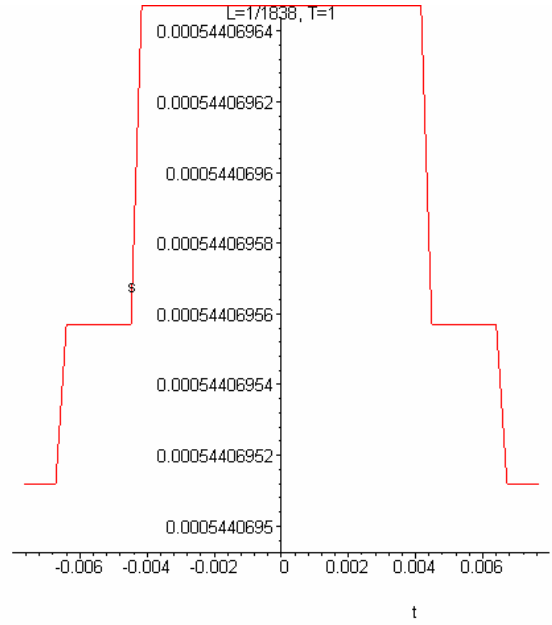

Fig.13. s- density $\tilde{\rho}_e$ in quantum soliton.

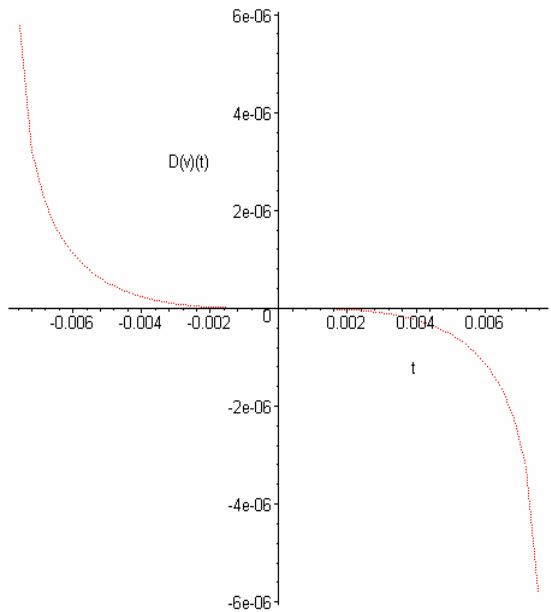

Fig. 14. The derivative of the self-consistent potential $\frac{\partial \tilde{\varphi}}{\partial \tilde{\xi}}$ in quantum soliton.

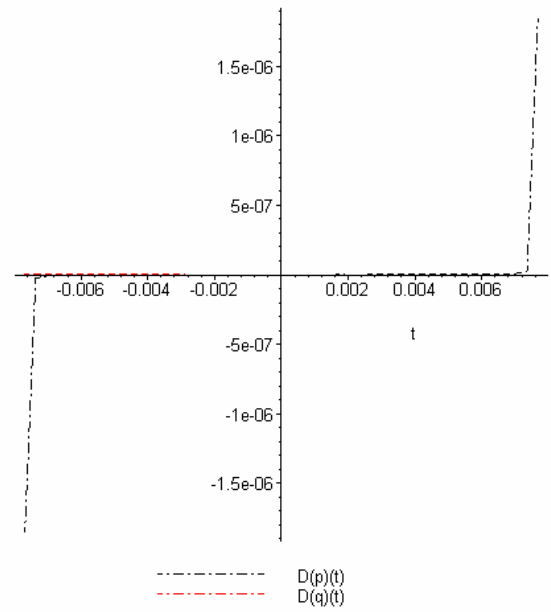

Fig. 15. The derivative of pressures $\frac{\partial \tilde{p}_i}{\partial \tilde{\xi}}$ and $\frac{\partial \tilde{p}_e}{\partial \tilde{\xi}}$ in quantum soliton.

Figures 12 – 15 remind the typical electron behavior in potential pit.
Consider the influence of changing of the rest non-local pressures $\tilde{p}_i(0)$, $\tilde{p}_e(0)$. Figures 16 -18 reflect the following Cauchy conditions:

```
v(0)=1,r(0)=1,s(0)=1/1838,u(0)=1,p(0)=1,q(0)=0.999,
D(v)(0)=0,D(r)(0)=0,D(s)(0)=0,D(u)(0)=0,D(p)(0)=0,D(q)(0)=0
```

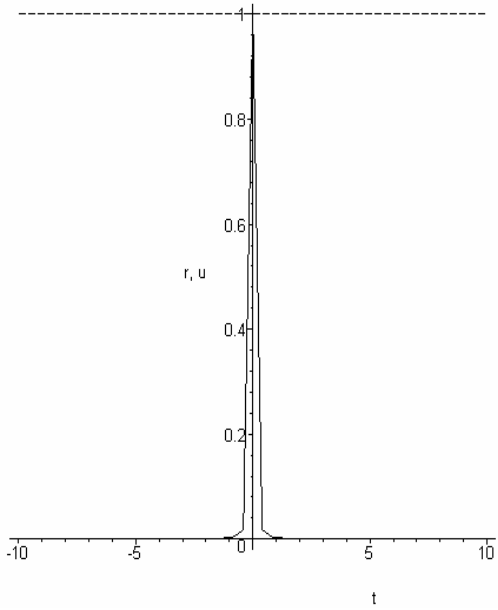 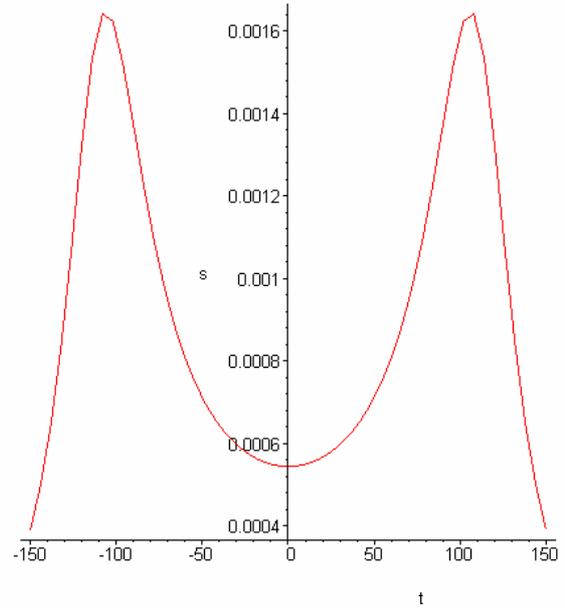

Fig.16. r- density $\tilde{\rho}_i$, u- velocity $\tilde{u}$ in quantum soliton.

Fig.17. s- density $\tilde{\rho}_e$ in quantum soliton.

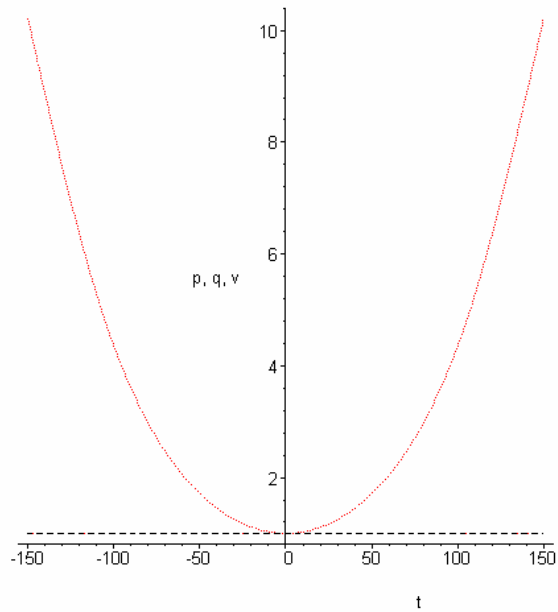

Fig. 18. v - self consistent potential $\tilde{\varphi}$ in quantum soliton and pressures $\tilde{p}_i$, $\tilde{p}_e$.

From Fig. 7- 11, 16 - 18 follows that increasing the difference $p_i(0) - p_e(0)$ lead to diminishing of the character domain occupied by soliton. The classical construction with the positive kernel and negative shell is existing if $\tilde{p}_i(0) > \tilde{p}_e(0)$. Moreover, in opposite case $\tilde{p}_i(0) < \tilde{p}_e(0)$ mathematical modeling leads to construction with negative kernel and positive shell for soliton. Let us demonstrate the possibility to calculate the soliton formations with very significant difference from the used scales. In following figures 19 – 21 this difference is in $10^5$ times more than the corresponding scales.

```
v(0)=10^5,r(0)=10^5,s(0)=(10^5)/1838,u(0)=1,p(0)=10^5,q(0)=0.95*
10^5,D(v)(0)=0,D(r)(0)=0,D(s)(0)=0,D(u)(0)=0,D(p)(0)=0,D(q)(0)=0
```

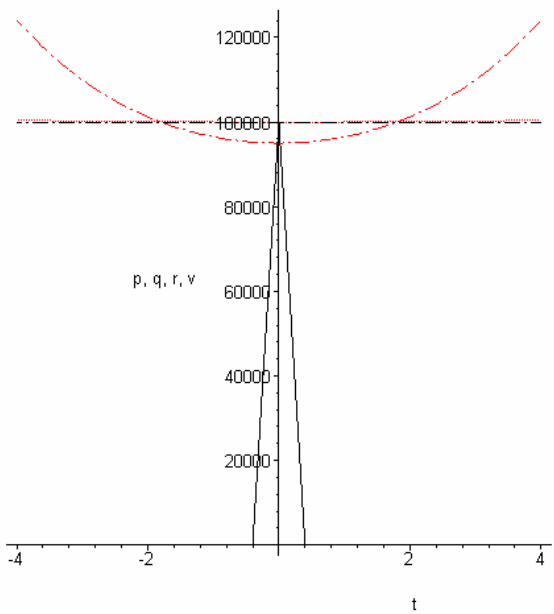 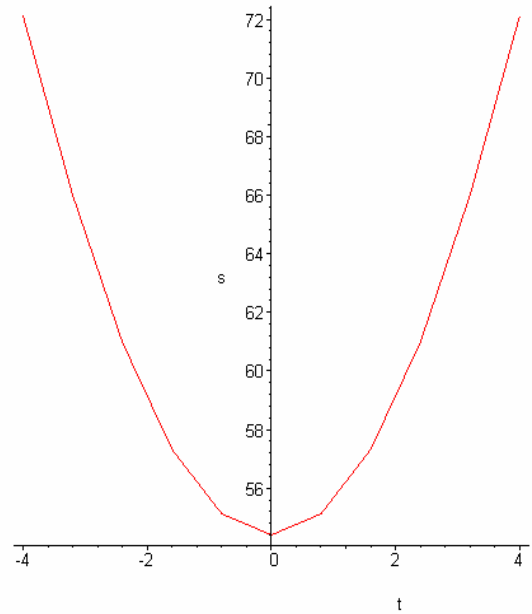

Fig.19. r- density $\tilde{\rho}_i$, p - pressure $\tilde{p}_i$, q - pressure $\tilde{p}_e$, Fig. 20. s- density $\tilde{\rho}_e$ in quantum soliton. v – self-consistent potential $\tilde{\varphi}$ in quantum soliton.

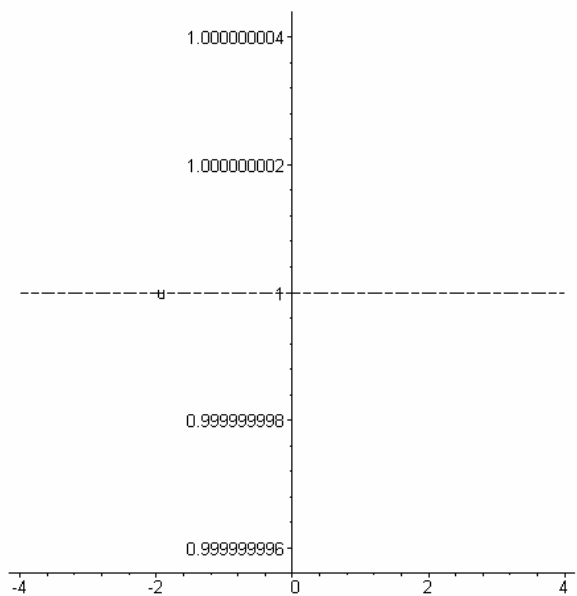

Fig.21. u- velocity $\tilde{u}$ in quantum soliton.

Let us demonstrate the influence of initial perturbation connected with gradient of potential, compare the case $\dfrac{\partial \tilde{\varphi}}{\partial \tilde{\xi}}(0) = 0$ (Fig. 7 – 11) after increasing this gradient in the following calculations to the value $\dfrac{\partial \tilde{\varphi}}{\partial \tilde{\xi}}(0) = 1$:

```
v(0)=1,r(0)=1,s(0)=1/1838,u(0)=1,p(0)=1,q(0)=.95,
D(v)(0)=1,D(r)(0)=0,D(s)(0)=0,D(u)(0)=0,D(p)(0)=0,D(q)(0)=0
```

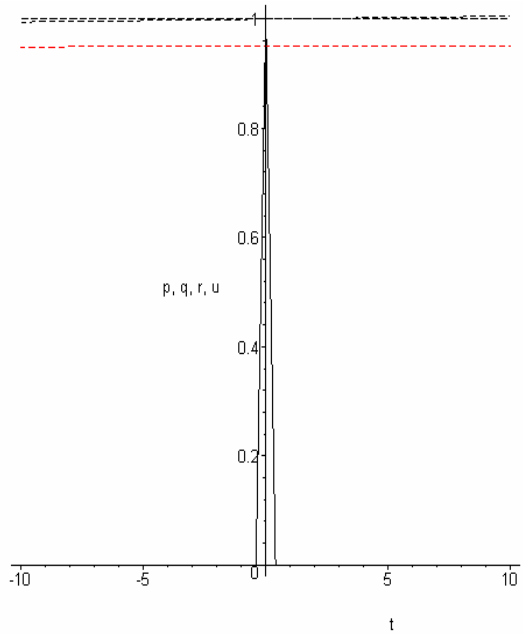 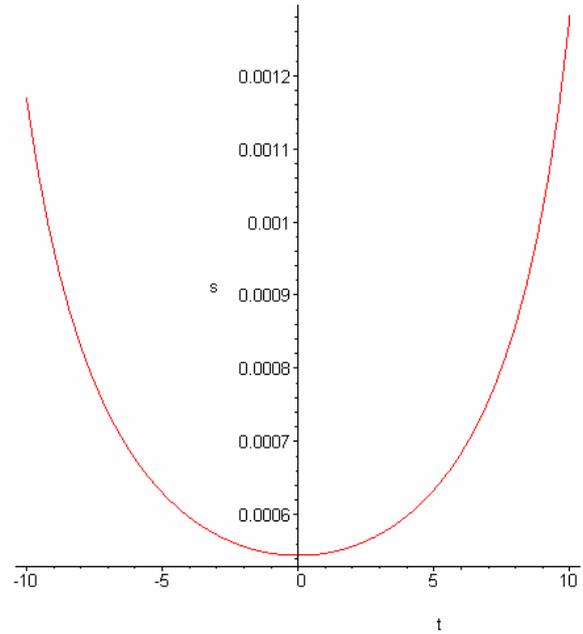

Fig.22. r- density $\tilde{\rho}_i$, u- velocity $\tilde{u}$, p - pressure $\tilde{p}_i$, q - pressure $\tilde{p}_e$ in quantum soliton.

Fig.23. s- density $\tilde{\rho}_s$ in quantum soliton

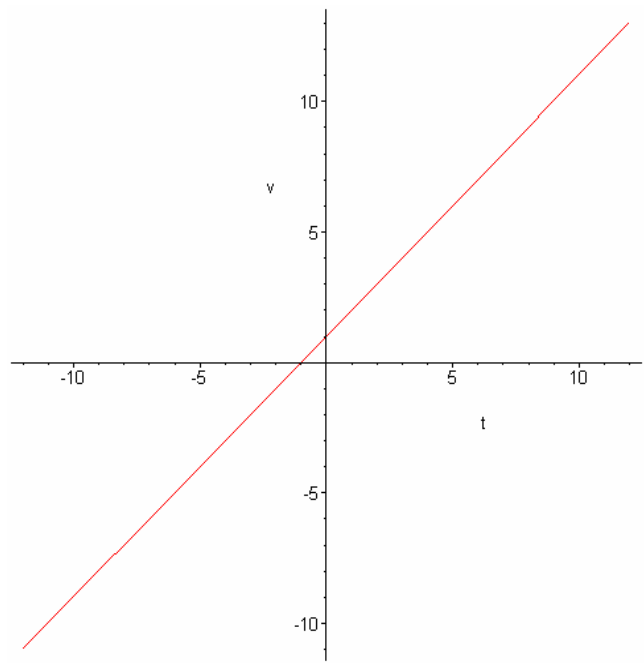

Fig. 24. v - self consistent potential $\tilde{\varphi}$ in quantum soliton

As we see from Figs. 7 -11, 22 – 24 this very significant changing of the initial gradient of potential does not provoke the destruction of soliton.

Important to notice that all elements of soliton are moving with the same self consistent constant velocity if initial perturbation $\tilde{u}(0)=1$ corresponds to phase velocity. The next calculation demonstrate the soliton destruction if the initial perturbation leads to another velocity of the object $(\tilde{u}(0)=0.5)$.

```
v(0)=1,r(0)=1,s(0)=1/1838,u(0)=0.5,p(0)=1,q(0)=0.95,
D(v)(0)=0,D(r)(0)=0,D(s)(0)=0,D(u)(0)=0,D(p)(0)=0,D(q)(0)=0
```

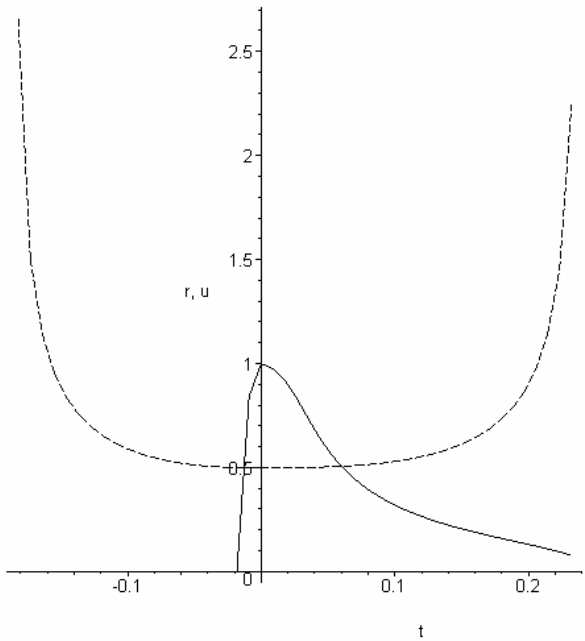
Fig.25. r- density $\tilde{\rho}_i$, u- velocity $\tilde{u}$.

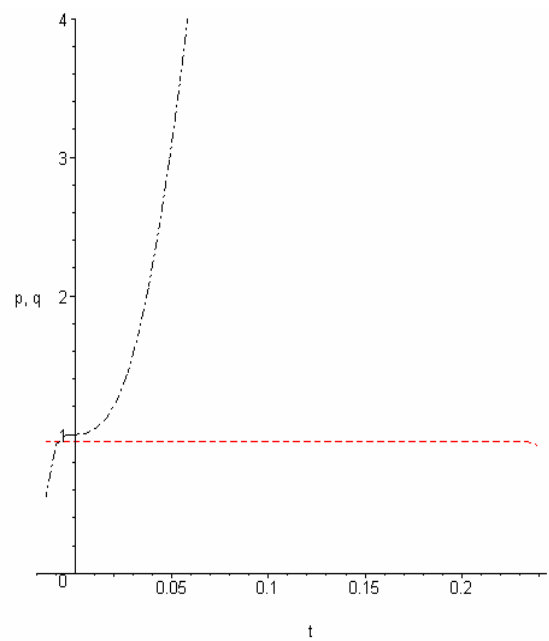
Fig.26. p - pressure $\tilde{p}_i$, q - pressure $\tilde{p}_e$.

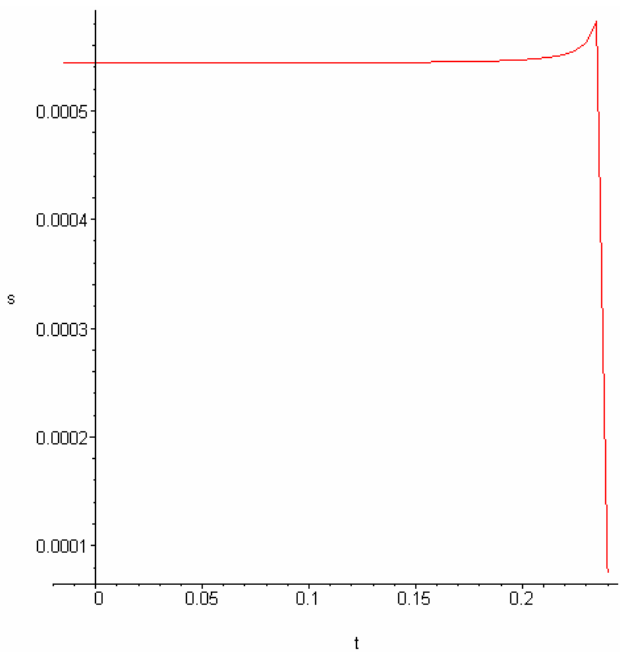
Fig.27. s- density $\tilde{\rho}_s$.

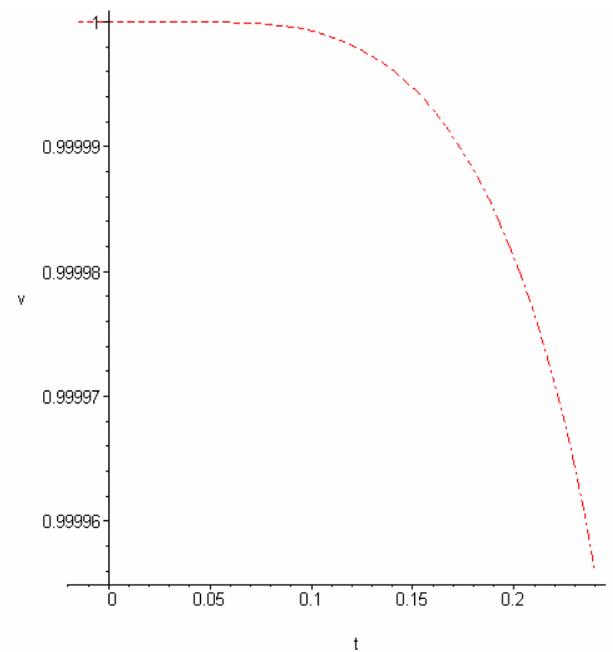
Fig.28. v - self consistent potential $\tilde{\varphi}$.

Let us consider now the situation when the soliton is catched by the external periodical longitudinal electric field $F_i^{(npot)} = \frac{eE}{m_i}\cos(kx - \omega t)$ for which phase velocity is equal $\omega/k = u_0$. In this case

$$\rho_i F_i^{(npot)} = \frac{eE}{m_i}\rho_0 \tilde{\rho}_i \cos\left[2\pi \frac{\hbar}{\lambda m_e u_0}\tilde{\xi}\right] \qquad (4.1)$$

And effective forces acting on positive and negative charges can be written as

$$\rho_i F_i^{(pot)} + \rho_i F_i^{(npot)} = -\frac{u_0^2}{x_0}\frac{m_e}{m_i}\rho_0 \tilde{\rho}_i \left[\frac{\partial \tilde{\varphi}}{\partial \tilde{\xi}} - \tilde{E}\cos\left(2\pi \frac{\hbar}{\lambda m_e u_0}\tilde{\xi}\right)\right], \qquad (4.2)$$

$$\rho_e F_e^{(pot)} + \rho_e F_e^{(npot)} = \frac{u_0^2}{x_0}\rho_0 \tilde{\rho}_e \left[\frac{\partial \tilde{\varphi}}{\partial \tilde{\xi}} - \tilde{E}\cos\left(2\pi \frac{\hbar}{\lambda m_e u_0}\tilde{\xi}\right)\right]. \qquad (4.3)$$

The expressions (4.2), (4.3) should be introduced in general system of quantum hydrodynamical equations (3.9) – (3.14). The amplitude $\tilde{E}$ and coefficient $\tilde{\Lambda} = 2\pi \frac{\hbar}{\lambda m_e u_0}$ are the parameters of calculations. Let us show the typical result of calculations in the external resonance electric field using the following Cauchy conditions with $\tilde{E}=1$, $\tilde{\Lambda}=1$ and $\tilde{p}_e(0) = \tilde{p}_i(0)$:

<span style="color:red">v(0)=1,r(0)=1,s(0)=1/1838,u(0)=1,p(0)=1,q(0)=1,
D(v)(0)=0,D(r)(0)=0,D(s)(0)=0,D(u)(0)=0,D(p)(0)=0,D(q)(0)=0</span>

Figures 29 – 32 reflect the results of calculations

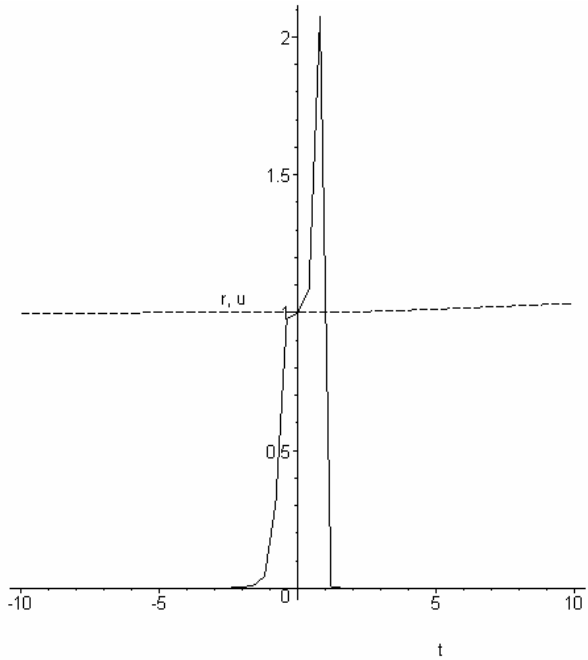
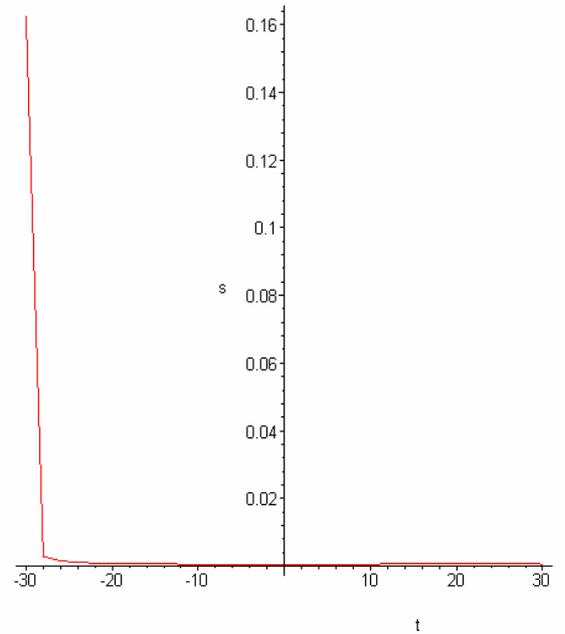

Fig.29. r- density $\tilde{\rho}_i$, u- velocity $\tilde{u}$.          Fig.30. s- density $\tilde{\rho}_s$.

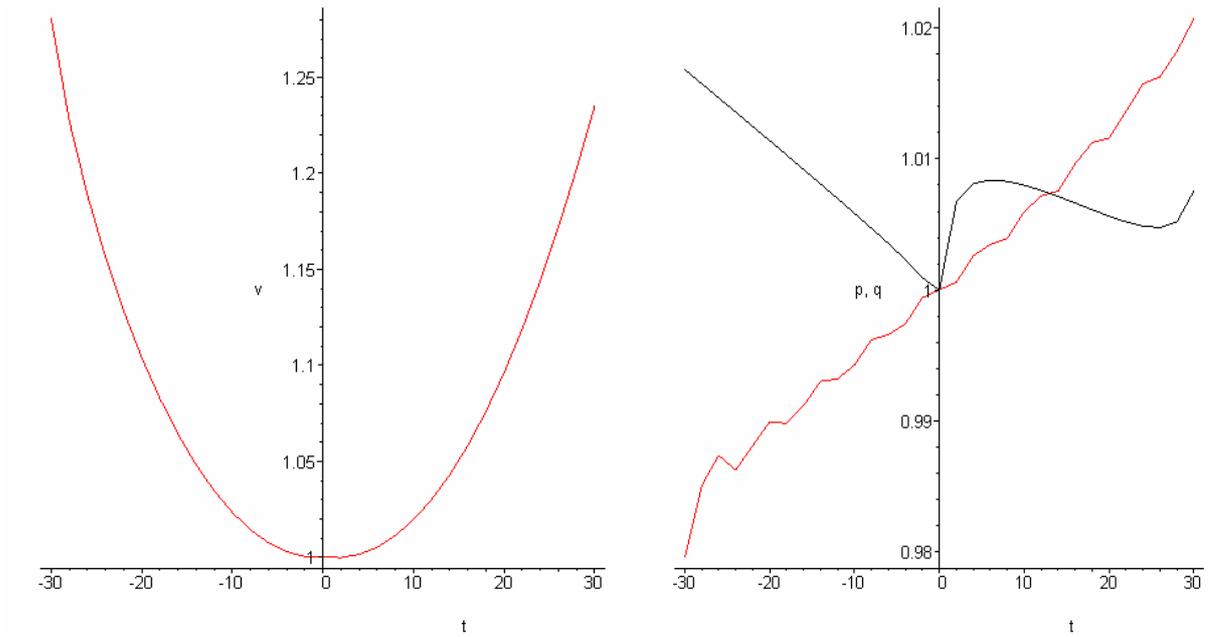

Fig.31. v - self consistent potential $\tilde{\varphi}$ ..   Fig.32. p - pressure $\tilde{p}_i$ (black line), $q$ -pressure $\tilde{p}_e$.

**Some remarks in conclusion.**

Quantum solitons are discovered with the help of generalized quantum hydrodynamics (GQH). The solitons have the character of quantum objects (with positive or negative shells) which reach stability as result of equalizing of corresponding pressure of the non-local origin and the self-consistent electric forces. These effects can be considered as explanation of the existence of lightning balls. If the initial rest pressures of non-local origin for the positive and negative components are equal each other but the quantum object is moving in the periodic resonance electric field, the mentioned quantum object is the stable soliton. In this case disappearing of the mentioned field leads to the blow up destruction of soliton.

The delivered theory demonstrates the great possibilities of the generalized quantum hydrodynamics in investigation of the quantum solitons, which creation and following evolution demands the non-steady 3D consideration in frame of GQH. The usual Schrödinger' quantum mechanics cannot be useful in this situation – at least but not at last – because Schrödinger – Madelung quantum theory does not contain the energy equation on principal.